\documentclass[prd,twocolumn,showpacs,preprintnumbers,amsmath,amssymb,superscriptaddress,floatfix,nofootinbib]{revtex4-2}

\usepackage{graphicx}
\usepackage{bm}
\usepackage{amsmath}
\usepackage{amsfonts}
\usepackage{amssymb}
\usepackage{color}
\usepackage{multirow}
\usepackage{subfigure}
\usepackage{txfonts}
\usepackage[colorlinks, citecolor=blue,anchorcolor=red,menucolor=red, linkcolor=red,filecolor=red,runcolor=red,urlcolor=blue,frenchlinks=red]{hyperref}

\begin{document}
\title{The roles of the $T_{c\bar{s}0}(2900)^0$ and $D_0^*(2300)$ in the process $B^-\to D_s^+K^-\pi^-$}

\author{Wen-Tao Lyu}
\affiliation{School of Physics and Microelectronics, Zhengzhou University, Zhengzhou, Henan 450001, China}

\author{Yun-He Lyu}
\affiliation{School of Physics and Microelectronics, Zhengzhou University,
             Zhengzhou, Henan 450001, China}\vspace{0.5cm}

\author{Man-Yu Duan}
\affiliation{School of Physics, Southeast University, Nanjing 210094, China}

\author{De-Min Li}\email{lidm@zzu.edu.cn}
\affiliation{School of Physics and Microelectronics, Zhengzhou University,
             Zhengzhou, Henan 450001, China}\vspace{0.5cm}

\author{Dian-Yong Chen}\email{chendy@seu.edu.cn}
\affiliation{School of Physics, Southeast University, Nanjing 210094, China}
\affiliation{Lanzhou Center for Theoretical Physics, Lanzhou University, Lanzhou 730000, China}

\author{En Wang}
\email{wangen@zzu.edu.cn}
\affiliation{School of Physics and Microelectronics, Zhengzhou University,
             Zhengzhou, Henan 450001, China}\vspace{0.5cm}
             \affiliation{Guangxi Key Laboratory of Nuclear Physics and Nuclear Technology, Guangxi Normal University, Guilin 541004, China}

\begin{abstract}
Motivated by the recent LHCb observations of $T_{c\bar{s}0}(2900)^0$ and $T_{c\bar{s}0}(2900)^{++}$ in the processes $B^0\to\bar{D}^0D_s^+\pi^-$ and $B^+\to D^-D_s^+\pi^+$, we have investigated the decay $B^-\to D_s^+K^-\pi^-$ by taking into account the contributions from the $S$-wave vector-vector interactions, and the $S$-wave $D^+_s K^-$ interactions. Our results show that the $D_s^+K^-$ invariant mass distribution has an enhancement structure near the threshold, associated with the $D^*_0(2300)$, which is in good agreement with the Belle measurements.  We have also predicted the $D^+_s\pi^-$ invariant mass distribution and the Dalitz plot, which show the significant signal of the $T_{c\bar{s}0}(2900)$. With the same formalism, the $D^-_sK^0_s$ invariant mass distribution of the process $B^0 \to D^-_sK^0_s\pi^+$ measured by Belle could be well reproduced, and the peak of $T_{c\bar{s}0}(2900)$ is expected to be observed around 2900~MeV in the $D^-_s\pi^+$ invariant mass distribution.
Our results could be tested by the Belle II and LHCb experiments in the future.

\end{abstract}

\date{\today}

\maketitle

\section{Introduction}\label{sec1}
Since the charmonium-like state $X(3872)$ was observed in the $\pi^+\pi^-J/\psi$ invariant mass distribution of the process $B^{\pm}\to K^{\pm}\pi^+\pi^-J/\psi$ by the Belle Collaboration in the year of 2003~\cite{Belle:2003nnu}, many candidates of the exotic states have been reported experimentally, and called many theoretical attentions, which largely deepens our understanding of the hadron spectra and hadron-hadron interactions~\cite{Chen:2016spr,Chen:2022asf,Guo:2017jvc,Oset:2016lyh}.

In the year of 2020, the LHCb Collaboration has observed two states $X_0(2900)$ and $X_1(2900)$ with masses about 2900~MeV in the $D^-K^+$ invariant mass distribution of the process $B^+\to D^+D^-K^+$~\cite{LHCb:2020bls,LHCb:2020pxc}, in agreement with the predictions of Ref.~\cite{Molina:2010tx}. 
As the open-flavor $\bar{c}\bar{s}ud$ states,  $X_0(2900)$ and $X_1(2900)$  have called many attentions, and there are several different theoretical interpretations about their nature, such as the compact tetraquark~\cite{He:2020jna,Karliner:2020vsi,Wang:2020prk,Yang:2021izl,Wang:2020xyc}, molecular structure interpretations~\cite{Wang:2021lwy,Chen:2020eyu,Lin:2022eau,Chen:2020aos,Huang:2020ptc,Liu:2020nil,Dai:2022qwh,Xiao:2020ltm,Hu:2020mxp}, or triangle singularity~\cite{Liu:2020orv}.

Recently, the LHCb Collaboration has reported two new states $T_{c\bar{s}0}(2900)^0$ and $T_{c\bar{s}0}(2900)^{++}$ in the $D_s^+\pi^-$ and $D_s^+\pi^+$ invariant mass distributions of the processes $B^0\to\bar{D}^0D_s^+\pi^-$ and $B^+\to D^-D_s^+\pi^+$ decays, respectively~\cite{LHCb:2022lzp,LHCb:2022sfr}, where the significance is found to be $8.0 \sigma$  for the $T_{c\bar{s}0}(2900)^0$ state and $6.5\sigma$ for the $T_{c\bar{s}0}(2900)^{++}$ state. Their masses and widths are determined as,
\begin{equation}
	\begin{aligned}
		& M_{T_{c \bar{s}0}(2900)^0}=(2892 \pm 14 \pm 15) ~\mathrm{MeV},\\
		& \Gamma_{T_{c \bar{s}0}(2900)^0}=(119 \pm 26 \pm 13)~ \mathrm{MeV}, \\
		& M_{T_{c \bar{s}0}(2900)^{++}}=(2921 \pm 17 \pm 20)~ \mathrm{MeV}, \\
		& \Gamma_{T_{c \bar{s}0}(2900)^{++}}=(137 \pm 32 \pm 17) ~\mathrm{MeV}, \\
		&
	\end{aligned}
\end{equation}
respectively. Their masses and widths are close to each other, which implies that  $T_{c\bar{s}0}(2900)^0$ and $T_{c\bar{s}0}(2900)^{++}$ with the flavor components $c\bar{s}\bar{u}d$ and $c\bar{s}u\bar{d}$ should be two of the isospin triplet.

Some interpretations for their structure are proposed theoretically such as the `genuine' tetraquark state~\cite{Lian:2023cgs,Meng:2023jqk,Yang:2023evp}, the molecular state~\cite{Agaev:2022duz,Duan:2023lcj}.
Since the $T_{c\bar{s}0}(2900)$ lies close to the thresholds of $D^*_s\rho$ and $D^*K^*$, the two-hadron continuum is expected to be of relevant for its existence, which make the $T_{c\bar{s}0}(2900)$ natural candidate for the molecular state~\cite{Guo:2017jvc,Matuschek:2020gqe}.
In Ref.~\cite{Agaev:2022duz}, 
the authors argue that $T_{c\bar{s}0}(2900)^{++}$ and $T_{c\bar{s}0}(2900)^0$ may be modelled as molecules $D_s^{*+}\rho^+$ and $D_s^{*+}\rho^-$, respectively, using two-point sum rule method. In addition, the $T_{c\bar{s}0}(2900)$ also can be considered as a virtual state created by the $D_s^*\rho$ and $D^*K^*$ interactions in coupled channels~\cite{Molina:2022jcd}, and the further analysis of the $D^*K^*$ interaction in a coupled-channel approach favors the $T_{c\bar{s}0}(2900)$ as a bound/virtual state\cite{Duan:2023lcj}. 
Thus, in this work we would like to study the production of $T_{c\bar{s}0}(2900)$ with the molecular scenario, which is expected to be checked by future experiments.

As we known, many candidates of the exotic states were observed in the decays of $B$ meson, and Belle and LHCb Collaborations have accumulated many events of $B$ mesons, which provides an important lab to study the hadron resonances~\cite{Chen:2022asf,Wang:2017mrt,Liu:2022dmm,Zhang:2020rqr,Dai:2018nmw,Li:2023nsw,Chen:2021erj}. For instance, we have proposed to search for the open-flavor tetraquark $T_{c\bar{s}0}(2900)^{++}$ in the process $B^+\to K^+D^+D^-$ by assuming $T_{c\bar{s}0}(2900)^{++}$ as a $D^{*+}K^{*+}$ molecular state in Ref.~\cite{Duan:2023qsg}. According 
to the Review of Particle Physics (RPP)~\cite{ParticleDataGroup:2022pth}, one can find that the branching fraction of the process $B^+ \to D^-_s \pi^+ K^+$ is 
$(1.80\pm 0.22)\times 10^{-4}$, in the same order of magnitude with the branching fraction of the process  $B^+\to D^- D^+ K^+$ 
[$(2.2\pm 0.7)\times 10^{-4}$], which implies that it is reasonable to search for the $T_{c\bar{s}0}(2900)$ in the process  $B^+\to D^-_s\pi^+ K^+$.

It should be pointed out that the processes $B^+\to D_s^- \pi^+ K^+$ and $B^-\to D_s^+ \pi^- K^-$ have been measured by the {\it BABAR}~\cite{BaBar:2007xlt} and Belle Collaborations~\cite{Belle:2009hlu}, and the $D_s^\pm K^\mp$ invariant mass distributions exhibit strong enhancements near the threshold. The same threshold enhancements have also appeared in the $D_s K$ invariant mass distribution in the processes $B^0\to D_s^-K^0_S\pi^+$ and  $B^+\to D_s^-K^+K^+$~\cite{Belle:2014agw}, which could be due to the contribution from the high pole of the $D^*_0(2300)$ with two-pole structure in the unitarized chiral perturbation theory~\cite{Albaladejo:2016lbb,Du:2019oki}\footnote{The state $D^*_0(2300)$ was denoted as $D_0^*(2400)$ in previous versions of RPP, and more discussions about this state can be found in the review `Heavy Non-$q\bar{q}$ mesons' of RPP~\cite{ParticleDataGroup:2022pth}.}.

Thus, in this work we would investigate the process $B^-\to D_s^+K^-\pi^-$ by taking into account the $S$-wave $D_s^{*+}\rho^-$ and $D^{*0}K^{*0}$ interactions, which will generate the resonance $T_{c\bar{s}0}(2900)$.   In addition, we will also consider the contribution from the $S$-wave pseudoscalar meson-pseudoscalar meson interactions within the unitary chiral approach, which will dynamically generate the resonance $D^*_0(2300)$.

This paper is organized as follows. In Sec.~\ref{sec2}, we present the theoretical formalism of the process $B^-\to D_s^+K^-\pi^-$. Numerical results and discussion are shown in Sec.~\ref{sec3}. Finally, we give a short summary in the last section.

 \section{Formalism}\label{sec2}
 
In this section, we will present the theoretical formalism of the process $B^-\to D_s^+K^-\pi^-$.  The reaction mechanism via the intermediate state $T_{c\bar{s}0}(2900)$ is given in Sec.~\ref{sec2a}, while the mechanism via the intermediate state $D^*_0(2300)$ is given in Sec.~\ref{sec2b}. Finally, we give the formalism of the invariant mass distributions of the process $B^-\to D_s^+K^-\pi^-$ in Sec.~\ref{sec2c}.

\begin{figure}[htbp]
 	\subfigure[]{
 	    \includegraphics[scale=0.65]{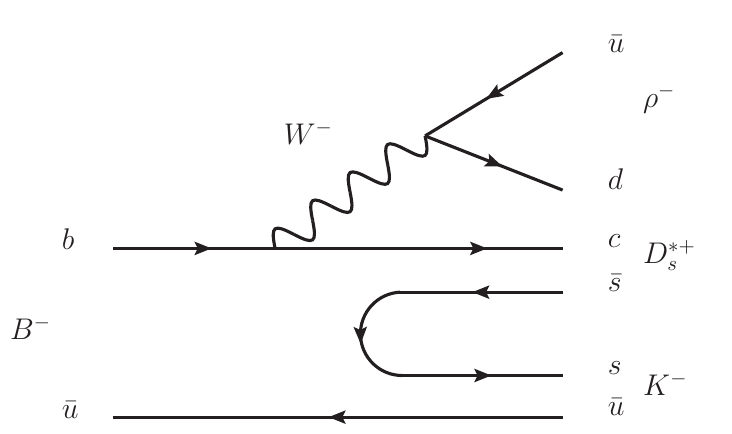}
 	}
 	
 	\subfigure[]{
 		\includegraphics[scale=0.65]{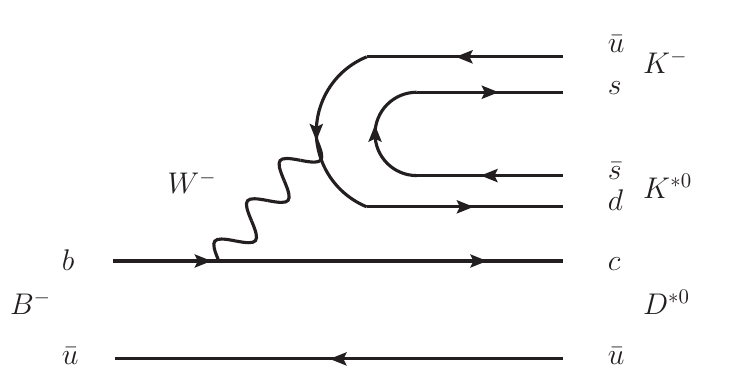}
 	}	
 	
 	\caption{ Quark level diagrams for the process $B^-\to \rho^-D_s^{*+}K^-$ via the $W^-$ external emission (a), and  the process $B^-\to K^{*0}D^{*0}K^-$ via the $W^-$ external emission (b).}\label{fig:Tcs-external}
\end{figure}

\subsection{The $T_{c\bar{s}0}(2900)$ role in $B^-\to D_s^+K^-\pi^-$} \label{sec2a}

\begin{figure}[htbp]
	\subfigure[]{
		\includegraphics[scale=0.65]{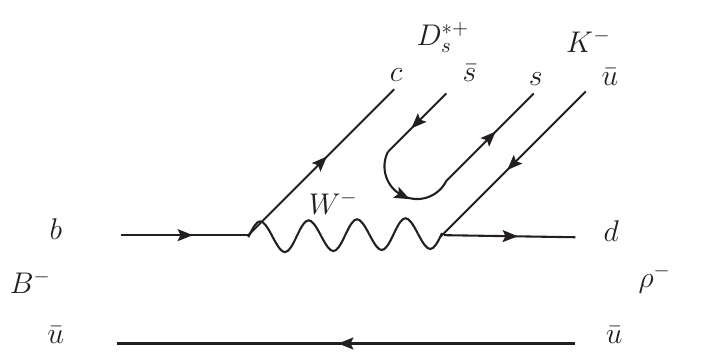}
	}
	
	\subfigure[]{
		\includegraphics[scale=0.65]{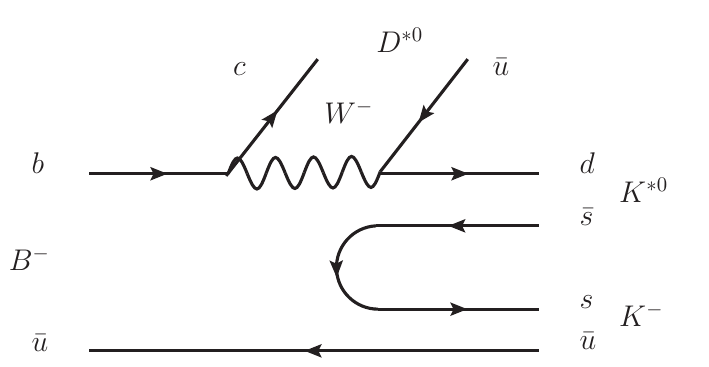}
	}
	\caption{ Quark level diagrams for the process $B^-\to \rho^-D_s^{*+}K^-$ process via the $W^-$ internal emission (a), and the process $B^-\to K^{*0}D^{*0}K^-$ via the $W^-$ internal emission (b).}\label{fig:Tcs-internal}
\end{figure}

Taking into account that $T_{c\bar{s}0}(2900)$ could be explained as the molecular state of the $D^*_s\rho$ and $D^*K^*$ interactions~\cite{Molina:2022jcd}, we first need to produce the states $D^{*+}_s\rho^- K^-$ and $D^{*0}K^{*0} K^-$ via the external $W^-$ emission mechanism and the internal $W^-$ emission mechanism, as depicted in Figs.~\ref{fig:Tcs-external} and \ref{fig:Tcs-internal}, respectively.

In analogy to Refs.~\cite{Wang:2017mrt,Wei:2021usz,Liu:2020ajv,Lu:2016roh,Wang:2015pcn,Chen:2015sxa},
as depicted in Fig.~\ref{fig:Tcs-external}(a), the $b$ quark of the initial $B^-$ meson weakly decays into a $c$ quark and a $W^-$ boson, then the $W^-$ boson decays into $\bar{u}d$ quarks. The $\bar{u}d$ pair from the $W^-$ boson will hadronize into $\rho^-$, while the $\bar{u}$ quark of the initial $B^-$ meson and the $c$ quark, together with the $\bar{s}s$ created from vacuum, hadronize into $K^-$ and $D_s^{*+}$. On the other hand, as shown in Fig.~\ref{fig:Tcs-external}(b),
the $\bar{u}d$ quarks from the $W^-$ boson, together with the $s\bar{s}$ created from vacuum, hadronize into $K^-$ and $K^{*0}$, while the $c$ quark and the $\bar{u}$ from the initial $B^-$ meson could hadronize into the $D^{*0}$.

In Fig.~\ref{fig:Tcs-internal}(a), the $c$ quark from the $B^-$ and the $\bar{u}$ from the $W^-$ boson, together with the $\bar{s}s$ created from vacuum, hadronize into $D_s^{*+}$ and $K^-$, while the $d$ quark from the $W^-$ boson and the $\bar{u}$ quark of the initial $B^-$ meson, hadronize into vector meson $\rho^-$. On the other hand, the $c$ quark from the $B^-$ and the $\bar{u}$ from the $W^-$ boson could also hadronize into a $D^{*0}$, while the $d$ quark from the $W^-$ boson and the $\bar{u}$ quark of the initial $B^-$ meson, together with the $\bar{s}s$ created from vacuum,  hadronize into  mesons $K^{*0}K^-$, as shown in Fig.~\ref{fig:Tcs-internal}(b). It should be pointed out that the mechanisms of Figs.~\ref{fig:Tcs-internal}(a) and \ref{fig:Tcs-internal}(b) are $1/N_c$ suppressed with respect to the ones of Fig.~\ref{fig:Tcs-external}.

\begin{figure}[htbp]
		\includegraphics[scale=0.65]{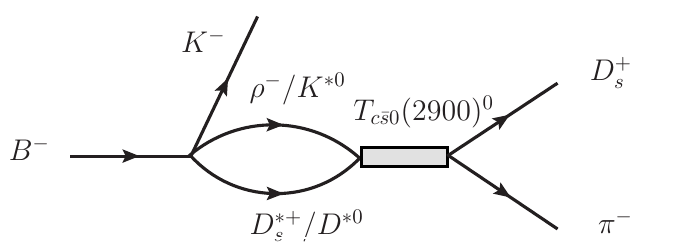}

	\caption{The mechanisms of the process $B^-\to D_s^+K^-\pi^-$ via the $S$-wave $\rho^-D_s^{*+}$ and $D^{*0}K^{*0}$ interactions.}\label{fig:Tcs2900_loop}
\end{figure}

Then, the $S$-wave interactions of  $\rho^-D_s^{*+}$ and $D^{*0}K^{*0}$ will give rise to the $T_{c\bar{s}0}(2900)^0$ state, which could decay into the final state $D_s^+\pi^-$, as depicted in Fig.~\ref{fig:Tcs2900_loop}. The transition amplitude for the process $B^-\to V_1 V_2 K^-$ ($V_1V_2= \rho^-D_s^{*+}$ or $D^{*0}K^{*0}$) could be written as,
\begin{equation}
\tilde{\mathcal{T}}= \mathcal{Q}(C+1) \vec\epsilon(V_1)\cdot \vec\epsilon(V_2),
\end{equation}
where $\vec\epsilon(V_1)$ and $\vec\epsilon(V_2)$ are the polarization vectors of $\rho^-D_s^{*+}$ (or $D^{*0}K^{*0}$), and we have the relation of $\sum_{\rm pol}\epsilon_i(V)\epsilon_j(V)=\delta_{ij}$. 
The constant $\mathcal{Q}$ includes all the dynamical factors of the weak decay of Fig.~\ref{fig:Tcs-internal}\footnote{The factor $\mathcal{Q}$ should weakly depend on the invariant mass of the vector-vector system, which does not influence the possible peak structure of $T_{c\bar{s}0}(2900)$. In addition, in this work we mainly focus on the intermediate resonances generated by the final state interactions, the parameter $\mathcal{Q}$ is assumed to be constant and independent of the final state interactions, as done in Refs.~\cite{Dai:2022qwh,Liu:2020orv,Dai:2018nmw,Wang:2017mrt,Liu:2022dmm}.}, while the factor $C=3$ corresponds to the relative weight of the $W^-$ external emission mechanism (Fig.~\ref{fig:Tcs-external}) with respect to the $W^-$ internal emission mechanism (Fig.~\ref{fig:Tcs-internal})~\cite{Duan:2020vye,Zhang:2022xpf,Wang:2021naf,Wang:2020pem}. Thus, one could easily obtain the expression of the $\mathcal{Q}$ as follows,
\begin{equation}\label{t6}
	\mathcal{Q}^2 \approx \frac{\Gamma_{B^-}{\mathcal B}(B^-\to D^{*0}K^{*0}K^-)}{\int \frac{3}{(2\pi)^3}\frac{(C+1)^2}{4M_{B^-}^2}p_{K^*}\tilde{p}_{K}{dM_{\rm inv}(D^{*0}K^{*0})}}.
\end{equation}

Since the branching fraction is measured to be $\mathcal{B}(B^-\to D^{*0}K^{*0}K^-)=(1.5\pm0.4)\times10^{-3}$~\cite{ParticleDataGroup:2022pth}, we could roughly estimate $\mathcal{Q}^2=1.71\times 10^{-13}$ neglecting the contributions from the possible intermediate resonances. 

By taking into account the contributions from the $S$-wave $\rho^-D_s^{*+}$ and $D^{*0}K^{*0}$ interactions of Fig.~\ref{fig:Tcs2900_loop}, the amplitude could be expressed as,
\begin{eqnarray}\label{ttcs}
		\mathcal{T}^{T_{c\bar{s}0}^0} &= &\mathcal{Q} \vec\epsilon(V_1)\cdot \vec\epsilon(V_2) \nonumber \\		
&&\times		(C+1)\left[G_{\rho^-D_s^{*+}}t_{\rho^-D_s^{*+}\to D_s^+\pi^-} \right. \nonumber \\
		&& \left.+ G_{D^{*0}K^{*0}}t_{D^{*0}K^{*0}\to D_s^+\pi^-}\right],
\end{eqnarray}
where $G_{\rho^-D_s^{*+}}$ and $G_{D^{*0}K^{*0}}$ are the loop functions of the coupled channels $\rho^-D_s^{*+}$ and $D^{*0}K^{*0}$, respectively, and $t_{\rho^-D_s^{*+}\to D_s^+\pi^-}$ and $t_{D^{*0}K^{*0}\to D_s^+\pi^-}$ are the transition amplitudes of $\rho^-D_s^{*+}\to D_s^+\pi^-$ and $D^{*0}K^{*0}\to D_s^+\pi^-$, respectively. Both of loop functions $G_i$ and transition amplitudes $t_{ij}$  are the functions of the $D_s^+\pi^-$ invariant mass $M_{D_s^+\pi^-}$. The two-meson loop function is given by,
\begin{equation}\label{G}
	G_i=i \int \frac{d^4 q}{(2 \pi)^4} \frac{1}{q^2-m_1^2+i \epsilon} \frac{1}{(P-q)^2-m_2^2+i \epsilon} ,
\end{equation}
where $m_1$ and $m_2$ are the mesons masses of the $i$-th coupled channel. $q$ is the four-momentum of the meson 1 in the center of mass frame, and $P$ is the four-momentum of the meson-meson system. In the present work, we use the dimensional regularization method as indicated in Refs.~\cite{Duan:2020vye,Duan:2023qsg}, and in this scheme, the two-meson loop function $G_i$ can be expressed as,
\begin{equation}
	\begin{aligned}
		G_i= & \frac{1}{16 \pi^2}\left\{a(\mu)+\ln \frac{m_1^2}{\mu^2}+\frac{s+m_2^2-m_1^2}{2 s} \ln \frac{m_2^2}{m_1^2}\right. \\
		& +\frac{|\vec{q}\,|}{\sqrt{s}}\left[\ln \left(s-\left(m_2^2-m_1^2\right)+2 |\vec{q}\,| \sqrt{s}\right)\right. \\
		& +\ln \left(s+\left(m_2^2-m_1^2\right)+2 |\vec{q}\,| \sqrt{s}\right) \\
		& -\ln \left(-s+\left(m_2^2-m_1^2\right)+2 |\vec{q}\,| \sqrt{s}\right) \\
		& \left.\left.-\ln \left(-s-\left(m_2^2-m_1^2\right)+2 |\vec{q}\,| \sqrt{s}\right)\right]\right\},
	\end{aligned}
\end{equation}
where $s=P^2=M_{\rm inv}^2(D_s^+\pi^-)$, and $\vec{q}$ is the three-momentum of the meson in the centre of mass frame, which reads,
\begin{equation}
	|\vec{q}\,|=\dfrac{\sqrt{[s-(m_1+m_2)^2][s-(m_1-m_2)^2]}}{2\sqrt{s}},
\end{equation}
here we take $\mu_{\rho^-D_s^{*+}}=\mu_{D^{*0}K^{*0}}=1500$~MeV, and $a_{\rho^-D_s^{*+}}=a_{D^{*0}K^{*0}}=-1.474$, which are the same as those used in the study of the $\rho^-D_s^{*+}$ interaction~\cite{Molina:2010tx,Molina:2022jcd} and the $D^{*0}K^{*0}$ interaction~\cite{Molina:2020hde,Molina:2022jcd,Duan:2023qsg}. And the transition amplitudes are given by,
\begin{equation}\label{t21}
	t_{\rho^-D_s^{*+}\to D_s^+\pi^-}=\dfrac{g_{T_{c\bar{s}0}^0,\rho^-D_s^{*+}}g_{T_{c\bar{s}0}^0,D_s^+\pi^-}}{M_{D_s^+\pi^-}^2-m_{T_{c\bar{s}0}^0}^2+im_{T_{c\bar{s}0}^0}\Gamma_{T_{c\bar{s}0}^0}},
\end{equation}
\begin{equation}\label{t22}
	t_{D^{*0}K^{*0}\to D_s^+\pi^-}=\dfrac{g_{T_{c\bar{s}0}^0,D^{*0}K^{*0}}g_{T_{c\bar{s}0}^0,D_s^+\pi^-}}{M_{D_s^+\pi^-}^2-m_{T_{c\bar{s}0}^0}^2+im_{T_{c\bar{s}0}^0}\Gamma_{T_{c\bar{s}0}^0}},
\end{equation}
where  the $m_{T_{c\bar{s}0}^0}$ and $\Gamma_{T_{c\bar{s}0}^0}$ are given by Refs.~\cite{LHCb:2022sfr,LHCb:2022lzp}. The constant $g_{T_{c\bar{s}0}^0,D^{*0}K^{*0}}$ corresponds to the coupling between $T_{c\bar{s}0}(2900)^0$ and its components $D^{*0}K^{*0}$, which could be related to the binding energy by the Weinberg compositeness criterion~\cite{Weinberg:1965zz,Baru:2003qq,Wu:2023fyh,Duan:2023qsg},
\begin{equation}\label{g_D^{*0}K^{*0}}
	g_{T_{c\bar{s}0}^0,D^{*}K^{*}}^2=16\pi(m_{D^*}+m_{K^*})^2\tilde{\lambda}^2\sqrt{\frac{2\Delta E}{\mu}},
\end{equation}
where $\tilde{\lambda}=1$ gives the probability to find the molecular component in the physical states. In this work we assume $T_{c\bar{s}0}(2900)^0$ as the $D^*K^*$ molecular state, and neglect possible $D^*_s\rho$ component, as done in Ref.~\cite{Duan:2023qsg}. 
$\Delta E=m_{D^*}+m_{K^*}-m_{T_{c\bar{s}0}}$ denotes the binding energy, and $\mu=m_{D^*}m_{K^*}/(m_{D^*}+m_{K^*})$ is the reduced mass. Here we obtain $g_{T_{c\bar{s}0}^0,D^{*0}K^{*0}}=8809$~MeV with Eq.~(\ref{g_D^{*0}K^{*0}}). 

Since the mass of $T_{c\bar{s}0}(2900)^0$ is larger than the thresholds of $D^*_s\rho$ and $D_s\pi$, the coupling constants $g_{T_{c\bar{s}0}^0,\rho^-D_s^{*+}}$ and $g_{T_{c\bar{s}0}^0,D_s^+\pi^-}$ could be obtained from the partial widths of $T_{c\bar{s}0}(2900)^0\to\rho^-D_s^{*+}$ and $T_{c\bar{s}0}(2900)^0\to D_s^{+}\pi^-$, respectively, which could be expressed as follows,
\begin{equation}\label{gg2}
	\Gamma_{T_{c\bar{s}0}^0\to \rho^-D_s^{*+}} =\frac{3}{8\pi}\frac{1}{m_{T_{c\bar{s}0}^0}^2}|g_{T_{c\bar{s}0}^0, \rho^-D_s^{*+}}|^2 |\vec{q}_{\rho}| ,
\end{equation}
\begin{equation}
	\Gamma_{T_{c\bar{s}0}^0\to D_s^{+}\pi^-} =\frac{1}{8\pi}\frac{1}{m_{T_{c\bar{s}0}^0}^2}|g_{T_{c\bar{s}0}^0,D_s^+\pi^-}|^2 |\vec{q}_{\pi^-}|, \label{eq:width_dspi}
\end{equation}
where
\begin{equation}
|\vec{q}_{\rho}|=\dfrac{\lambda^{1/2}(m_{T_{c\bar{s}0}^0}^2,m_{D_s^{*+}}^2,m_{\rho^-}^2)}{2m_{T_{c\bar{s}0}^0}},
\end{equation}
\begin{equation}
|\vec{q}_{\pi^-}|=\dfrac{\lambda^{1/2}(m_{T_{c\bar{s}0}^0}^2,m_{D_s^{+}}^2,m_{\pi^-}^2)}{2m_{T_{c\bar{s}0}^0}}.
\end{equation}
with the K$\ddot{a}$llen function $\lambda(x,y,z)=x^2+y^2+z^2-2xy-2yz-2zx$. 
In Ref.~\cite{Yue:2022mnf}, the partial widths of decay modes $\rho^-D_s^{*+}$ and $D_s^+\pi^-$ were estimated to be ($2.96 \sim 5.3$)~MeV and ($0.55 \sim 8.35$)~MeV, respectively. In the present work, we take the center values of the decay widths $\Gamma_{T_{c\bar{s}0}^0\to \rho^-D_s^{*+}}=4.13$~MeV and $\Gamma_{T_{c\bar{s}0}^0\to D_s^{+}\pi^-}=4.45$~MeV~\cite{Yue:2022mnf}, from which we can obtain the coupling constants $g_{T_{c\bar{s}0}^0,\rho^-D_s^{*+}}=2007$~MeV and  $g_{T_{c\bar{s}0}^0,D_s^+\pi^-}=1104$~MeV, respectively. One can find that the coupling of  $g_{T^0_{c\bar{s}0}D^{*0}K^{*0}}$ is larger than two other couplings, which implies that the $D^*K^*$ component plays the dominant role for $T_{c\bar{s}0}(2900)^0$.

\subsection{The $D^*_{0}(2300)$ role in $B^-\to D_s^+K^-\pi^-$} \label{sec2b}

\begin{figure}[htbp]
	
		\subfigure[]{
		\includegraphics[scale=0.65]{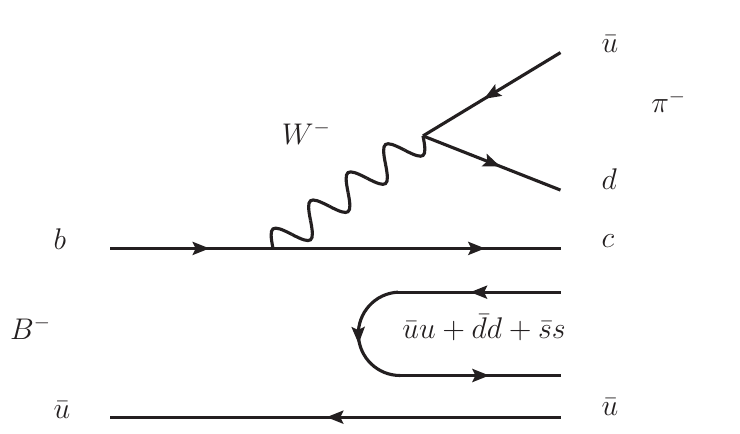}
	}
	\subfigure[]{
		\includegraphics[scale=0.65]{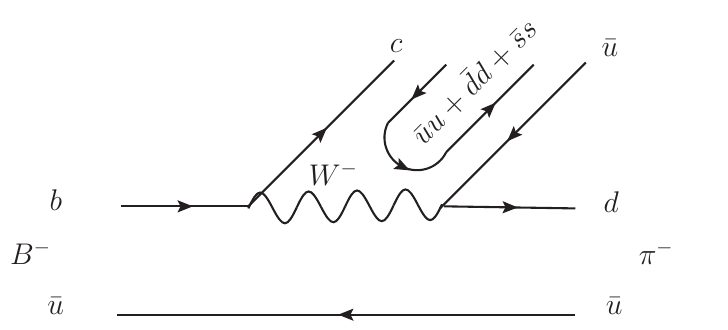}
	}
	
	\caption{Quark level diagrams for the $B^-\to \pi^- c(\bar{u}u+\bar{d}d+\bar{s}s)\bar{u}$ process. (a) The $W^-$ external emission, (b) the $W^-$ internal emission.}\label{fig:D(2300)-quark}
\end{figure}

In this subsection, we will consider the contribution from the $S$-wave $D_s^+ K^-$ final state interaction. As shown in Fig.~\ref{fig:D(2300)-quark}, the $b$ quark of the initial $B^-$ weakly decays into a $c$ quark and a $W^-$ boson, then the $W^-$ boson subsequently decays into a $\bar{u}d$ quark pair, which will hadronize into a $\pi^-$ meson.  The $c$ quark from the $B^-$ decay and the $\bar{u}$ quark of the initial $B^-$, together with the quark pair $\bar{q}q=\bar{u}u+\bar{d}d+\bar{s}s$ created from the vacuum with the quantum numbers $J^{PC}=0^{++}$, hadronize into hadrons pairs, as follows,
\begin{equation}
	\sum_ic(\bar{u}u+\bar{d}d+\bar{s}s)\bar{u} = \sum_{i=1}^3P_{4i}P_{i1},
\end{equation}
where $i=1,2,3$ correspond to the $u$, $d$, and $s$ quarks, respectively, and $P$ is the U(4) matrix of the pseudoscalar mesons,
\begin{equation}
	P=\left(\begin{array}{cccc}
		\frac{\pi^0}{\sqrt{2}} +\frac{\eta}{\sqrt{3}}  +\frac{\eta'}{\sqrt{6}} & \pi^{+} & K^{+} & \bar{D}^0 \\
		\pi^{-} & -\frac{\pi^0}{\sqrt{2}} +\frac{\eta}{\sqrt{3}} +\frac{\eta'}{\sqrt{6}} & K^0 & D^{-} \\
		K^{-} & \bar{K}^0 & -\frac{\eta}{\sqrt{3}} +\frac{2\eta'}{\sqrt{6}} & D_s^{-} \\
		D^0 & D^{+} & D_s^{+} & \eta_c
	\end{array}\right) ,
\end{equation}
where we have taken the  approximate $\eta-\eta'$ mixing from Ref.~\cite{Bramon:1992kr}. \footnote{According to PRR~\cite{ParticleDataGroup:2022pth}, the mixing angle is between $-10^\circ$ and $-20^\circ$, and a  recent Lattice calculations support the value $\theta_P=-14.1^\circ \pm 2.8^\circ$ by
 reproducing the masses of the $\eta$ and $\eta'$~\cite{Christ:2010dd}. In this work, we adopt the mixing from Ref.~\cite{Bramon:1992kr},
  $\eta\sim \frac{1}{\sqrt{3}}(u\bar{u}+d\bar{d}-s\bar{s})$ and 
$\eta' \sim \frac{1}{\sqrt{6}}(u\bar{u}+d\bar{d}+2s\bar{s})$.}

Then, we could have all the possible pseudoscalar-
pseudoscalar pairs after the hadronization,
\begin{equation}\label{H}
 H=\pi^-\left(\dfrac{1}{\sqrt{2}}D^0\pi^0+\dfrac{1}{\sqrt{3}}D^0\eta+D^+\pi^-+D_s^+K^-\right).
\end{equation}

With the isospin multiplets of ($D^+, -D^0$), ($\bar{D}^0, D^-$), and ($-\pi^+, \pi^0, \pi^-$), we have,
\begin{equation}
	\begin{aligned}
	D^0\pi^0&=-\left|\frac{1}{2},-\frac{1}{2}\right> \left|1,0\right> \\
	&=-\sqrt{\frac{2}{3}}\left|\frac{3}{2},-\frac{1}{2}\right>+\sqrt{\frac{1}{3}}\left|\frac{1}{2},-\frac{1}{2}\right>,
	\end{aligned}
\end{equation}
\begin{equation}
	\begin{aligned}
		D^+\pi^-&=\left|\frac{1}{2},\frac{1}{2}\right>\left|1,-1\right> \\
		&=\sqrt{\frac{1}{3}}\left|\frac{3}{2},-\frac{1}{2}\right>+\sqrt{\frac{2}{3}}\left|\frac{1}{2},-\frac{1}{2}\right>,
	\end{aligned}
\end{equation}
\begin{equation}
	\begin{aligned}
		\dfrac{1}{\sqrt{2}}D^0\pi^0+D^+\pi^-&=\left(-\sqrt{\frac{2}{3}}\cdot\sqrt{\frac{1}{2}}+\sqrt{\frac{1}{3}}\right)|D\pi>^{I=\frac{3}{2}} \\
		&+\left(\sqrt{\frac{1}{3}}\cdot\sqrt{\frac{1}{2}}+\sqrt{\frac{2}{3}}\right)|D\pi>^{I=\frac{1}{2}}, \\
		&=\sqrt{\frac{3}{2}}|D\pi>^{I=\frac{1}{2}}.
	\end{aligned}
\end{equation}

\begin{figure}[htbp]
	\subfigure[]{
		\includegraphics[scale=0.65]{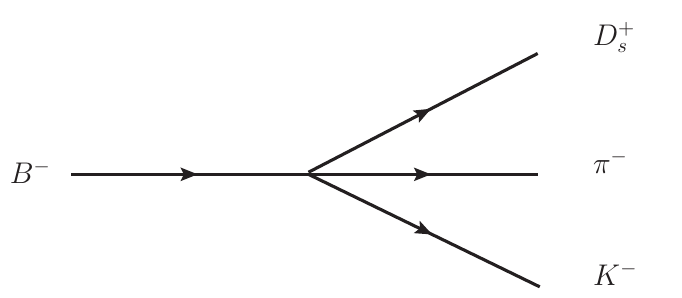}
	}

	\subfigure[]{
		\includegraphics[scale=0.65]{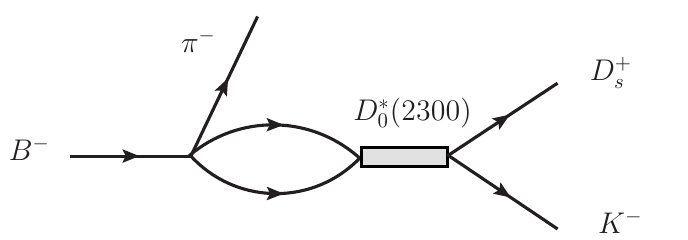}
	}
	\caption{The mechanisms of the $B^-\to D_s^+K^-\pi^-$ decay. (a) Tree diagram, (b) the mechanism of the $S$-wave meson-meson interactions.}\label{fig:D(2300)_hadron}
\end{figure}

In the isospin basis, we can obtain the $D\pi(I=1/2)$, $D\eta(I=1/2)$, and $D_s\bar{K}(I=1/2)$ channels
\begin{equation}\label{H}
 H=\pi^-\left({\sqrt{\frac{3}{2}}}D\pi+\dfrac{1}{\sqrt{3}}D^0\eta+D_s^+K^-\right) .
\end{equation}

Then, the process $B^-\to D_s^+K^-\pi^-$ decay  could happen via the tree diagram of Fig.~\ref{fig:D(2300)_hadron}(a), and the $S$-wave meson-meson interaction of Fig.~\ref{fig:D(2300)_hadron}(b), and the amplitude could be expressed as,
\begin{equation}\label{t-D2300}
	\begin{aligned}
		\mathcal{T}^{D_0^*(2300)} &= \mathcal{Q}^{\prime}(C+1)(h_{D_s\bar{K}} + \sum_ih_iG_it_{i\to D_s\bar{K}}) \\
		&=\mathcal{T}^{\rm tree}+\mathcal{T}^{S},
	\end{aligned}
\end{equation}
where the constant $\mathcal{Q}^{\prime}$ includes all the dynamical factors of the weak decay, and  $i=1,2,3$ correspond to the $D\pi$, $D\eta$, $D_s\bar{K}$, respectively,
\begin{equation}
  		h_{D\pi} =\sqrt{\frac{3}{2}}, \quad h_{D\eta}=\sqrt{\frac{1}{3}}, \quad h_{D_s\bar{K}}=1.
\end{equation}

The factor $C=3$ corresponds to the relative weight of the $W^-$ external emission mechanism [Fig.~\ref{fig:D(2300)-quark}(a)] with respect to the $W^-$ internal emission mechanism [Fig.~\ref{fig:D(2300)-quark}(b)]. With the amplitude of Eq.~(\ref{t-D2300}), the $\mathcal{Q}^{\prime}$ is given by,
\begin{equation}\label{t7}
	\begin{aligned}
	&\Gamma_{B^-}\mathcal{B}(B^-\to D_s^+K^-\pi^-)  \\
	&=\mathcal{Q}^{\prime 2}\int \frac{1}{(2\pi)^3}\frac{(C+1)^2}{4M_{B^-}^2}p_{\pi}\tilde{p}_K \\
	&\times|h_{D_s\bar{K}} + \sum_ih_iG_it_{i\to D_s\bar{K}}|^2{dM_{\rm inv}(D_s^+K^-)}.
    \end{aligned}
\end{equation}

According   to the experimental measurements of $\mathcal{B}(B^-\to D_s^+K^-\pi^-)=(1.80\pm0.22)\times10^{-4}$~\cite{ParticleDataGroup:2022pth}, we could roughly estimate $\mathcal{Q}^{\prime 2}=1.81\times10^{-12}$.

The $G_i$ in Eq.~(\ref{t-D2300}) is the loop function of the meson-meson system, and $t_{i\to D_s\bar{K}}$ are the scattering matrices of the coupled channels. The transition amplitude of $t_{i\to D_s\bar{K}}$ is obtained by solving the Bethe-Salpeter equation,
\begin{equation}\label{BS}
	T=[1-VG]^{-1}V.
\end{equation}

The transition potential $V_{ij}$ is taken from Refs.~\cite{Liu:2012zya,Montana:2020vjg},
\begin{equation}\label{V}
	\begin{aligned}
	V_{ij} &=\frac{1}{4f_{\pi}^2}C_{ij}\left(s-u\right) \\
	&= \frac{C_{ij}}{4f_{\pi}^2}\left(2s-m_2^2-m_4^2-2E_1E_3\right),
	\end{aligned}
\end{equation}
where the coefficients are given as,
\begin{equation}
	C_{ij}=\left(\begin{array}{ccc}
		-2 & 0 & -\sqrt{\frac{3}{2}} \\
		0 & 0 & -\sqrt{\frac{3}{2}} \\
		-\sqrt{\frac{3}{2}} & -\sqrt{\frac{3}{2}} & -1
	\end{array}\right) ,
\end{equation}
with $f_{\pi}=93$~MeV. The loop function $G_i$ in Eq.~(\ref{BS}) is given by the dimensional regularization method, as shown by Eq.~(\ref{G}), and we take $\mu=1000$~MeV and $a=-1.88$~\cite{Liu:2012zya}.

\subsection{Invariant Mass Distribution} \label{sec2c}

With above the formalism, one can write down the invariant mass distribution for the $B^-\to D_s^+K^-\pi^-$,
\begin{equation}\label{dwidth}
	\dfrac{d^2\Gamma}{dM_{D_s^+K^-}dM_{D_s^+\pi^-}} = \frac{1}{(2\pi)^3}\dfrac{2M_{D_s^+K^-}2M_{D_s^+\pi^-}}{32M_{B^-}^3}|\mathcal{T}^{\text {total }}|^2,
\end{equation}
where the modulus squared of the total amplitude is,
\begin{equation}\label{ttotal-interference}
	\left|\mathcal{T}^{\text {total }}\right|^2=\left|\mathcal{T}^{T_{c\bar{s}0}^0}e^{i\phi}+\mathcal{T}^{D_0^*(2300)}\right|^2,
\end{equation}
with a phase $\phi$ between two terms.
For a given value of invariant mass $M_{12}$, the range of invariant mass $M_{23}$ is determined by~\cite{ParticleDataGroup:2022pth},
\begin{align}
	&\left(m_{23}^2\right)_{\min}=\left(E_2^*+E_3^*\right)^2-\left(\sqrt{E_2^{* 2}-m_2^2}+\sqrt{E_3^{* 2}-m_3^2}\right)^2, \nonumber\\
	&\left(m_{23}^2\right)_{\max}=\left(E_2^*+E_3^*\right)^2-\left(\sqrt{E_2^{* 2}-m_2^2}-\sqrt{E_3^{* 2}-m_3^2}\right)^2, 
\end{align}
where $E_2^{*}$ and $E_3^{*}$ are the energies of particles 2 and 3 in the $M_{12}$ rest frame. $E_2^{*}$ and $E_3^{*}$ are written as,
\begin{align}
	&E_2^{*}=\dfrac{M_{12}^2-m_1^2+m_2^2}{2M_{12}}, \nonumber\\
	&E_3^{*}=\dfrac{M_{B^-}^2-M_{12}^2+m_3^2}{2M_{12}},
\end{align}
where $m_1$, $m_2$, and $m_3$ are the masses of particles 1, 2, and 3, respectively. All the masses and widths of the particles are taken from the RPP~\cite{ParticleDataGroup:2022pth}.

\section{ Numerical results }\label{sec3}

\begin{figure}[htbp]
	
	\includegraphics[scale=0.65]{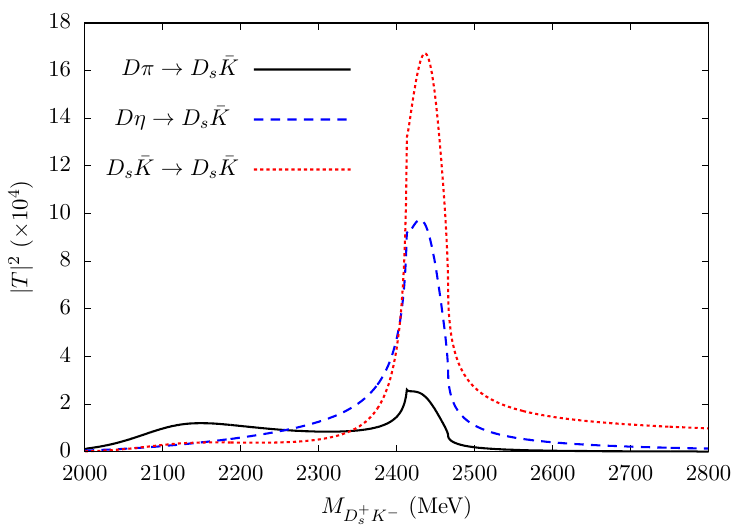}

	\caption{Modulus squared of the transition amplitudes $t_{i\to D_s\bar{K}}$ in $S$-wave.}\label{fig:square-T}
\end{figure}

\begin{figure}[htbp]
	
	\includegraphics[scale=0.65]{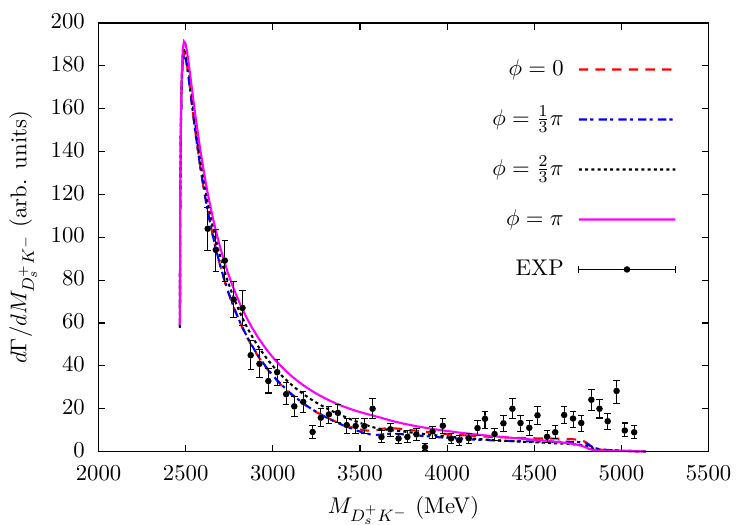}
	
	\caption{The $D_s^+K^-$ invariant mass distribution of the process $B^-\to D_s^+K^-\pi^-$ with the interference phase $\phi=0$, $\frac{1}{3}\pi$, $\frac{2}{3}\pi$, and $\pi$. Belle data have been rescaled for comparison~\cite{Belle:2009hlu}.}\label{fig:dwidth-DsK-interference}
\end{figure}

\begin{figure}[htbp]
	
	\includegraphics[scale=0.65]{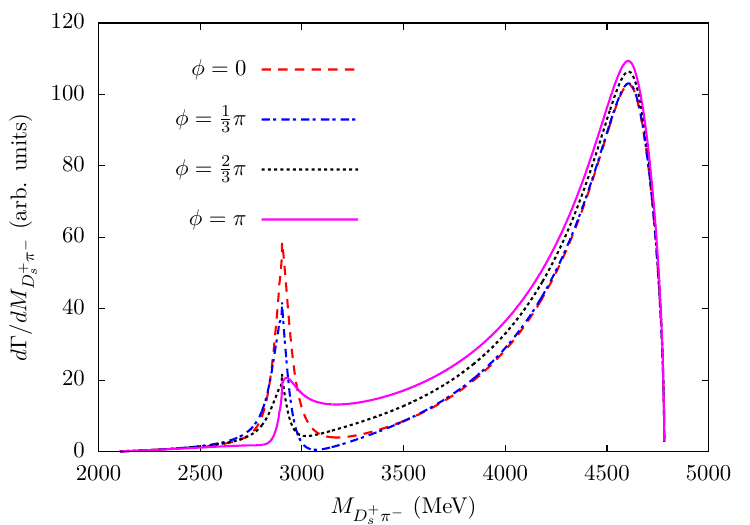}
	
	\caption{The $D_s^+\pi^-$ invariant mass distribution of the process $B^-\to D_s^+K^-\pi^-$  with the interference phase $\phi=0$, $\frac{1}{3}\pi$, $\frac{2}{3}\pi$, and $\pi$. }\label{fig:dwidth-dspi-interference}
\end{figure}

\begin{figure*}[htbp]
	\subfigure[]{
		\includegraphics[scale=0.65]{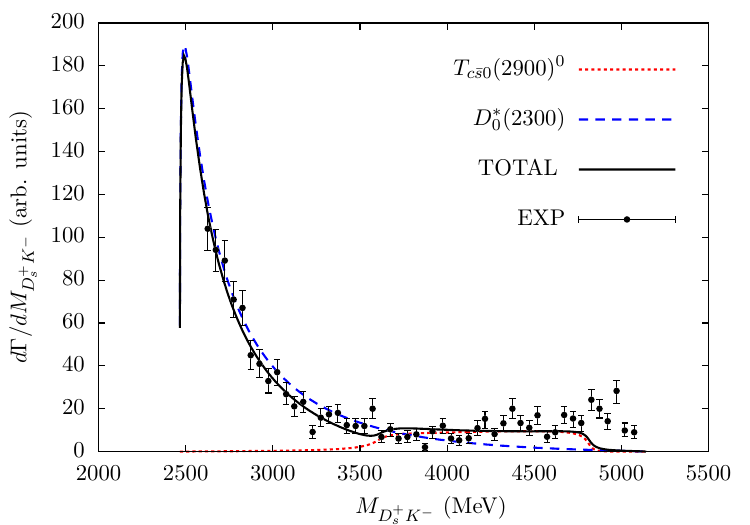}
	}	
	\subfigure[]{
		\includegraphics[scale=0.65]{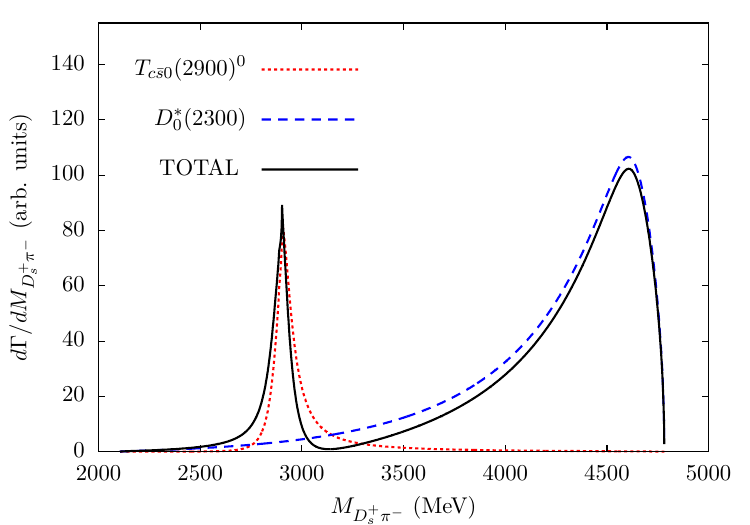}
	}
	\caption{The $D_s^+K^-$ (a) and $D_s^+\pi^-$ (b)  invariant mass distributions of the process $B^-\to D_s^+K^-\pi^-$ with the fitted parameters $\Gamma_{T_{c\bar{s}0}^0\to D_s^{+}\pi^-}=10.45$~MeV and $\phi=0.35\pi$. The Belle data are taken from the Ref.~\cite{Belle:2009hlu}.}\label{fig:dwidth-interference-fit}
\end{figure*}

\begin{figure}[htbp]
	\includegraphics[scale=0.85]{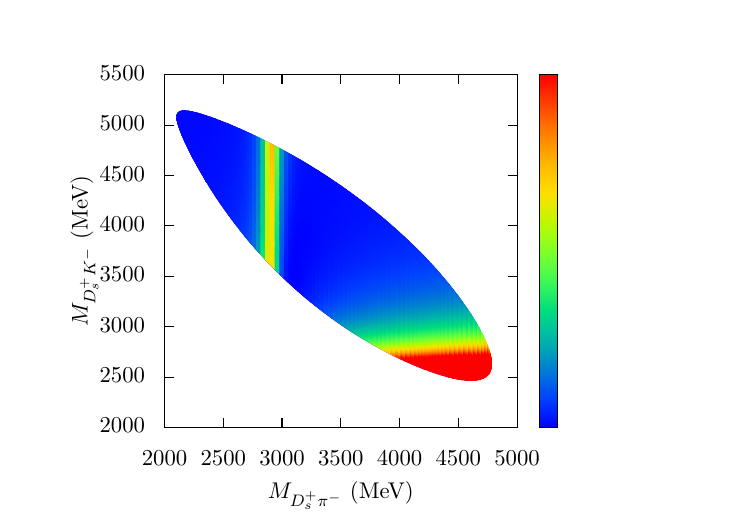}	
	\caption{The Dalitz plot of ``$M_{D_s^+\pi^-}$" vs. ``$M_{D_s^+K^-}$" for the process $B^-\to D_s^+K^-\pi^-$ with the fitted parameters $\Gamma_{T_{c\bar{s}0}^0\to D_s^{+}\pi^-}=10.45$~MeV and $\phi=0.35\pi$.}\label{fig:Dalitz-dwidth}
\end{figure}

\begin{figure*}[htbp]
	\subfigure[]{
		\includegraphics[scale=0.65]{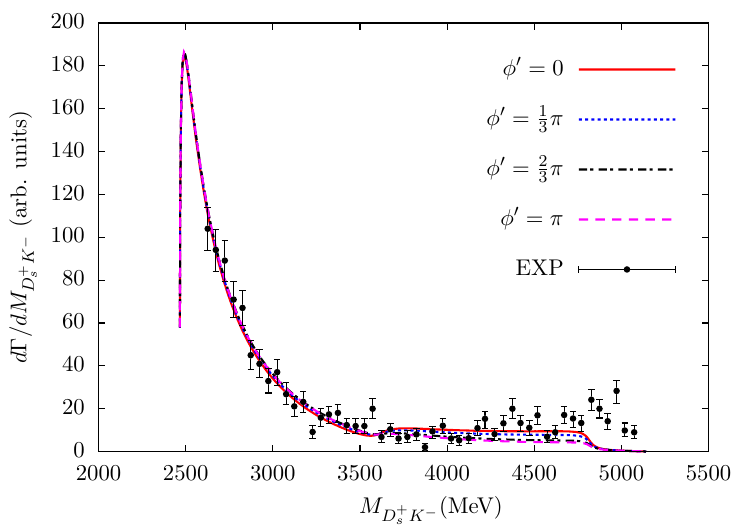}
	}	
	\subfigure[]{
		\includegraphics[scale=0.65]{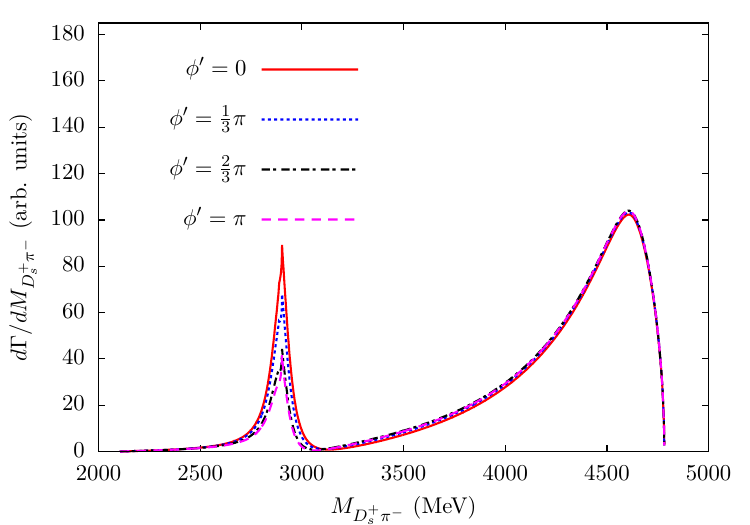}
	}
	\caption{The $D_s^+K^-$ (a) and $D_s^+\pi^-$ (b)  invariant mass distributions of the process $B^-\to D_s^+K^-\pi^-$ with the fitted parameters $\Gamma_{T_{c\bar{s}0}^0\to D_s^{+}\pi^-}=10.45$~MeV and $\phi=0.35\pi$, where the phase $\phi'$ is taken to be $\phi'=0$, $\frac{1}{3}\pi$, $\frac{2}{3}\pi$, and $\pi$, respectively. The Belle data are taken from the Ref.~\cite{Belle:2009hlu}. }\label{fig:dwidth-phiprime}
\end{figure*}

\begin{figure*}[htbp]
	\subfigure[]{
		\includegraphics[scale=0.65]{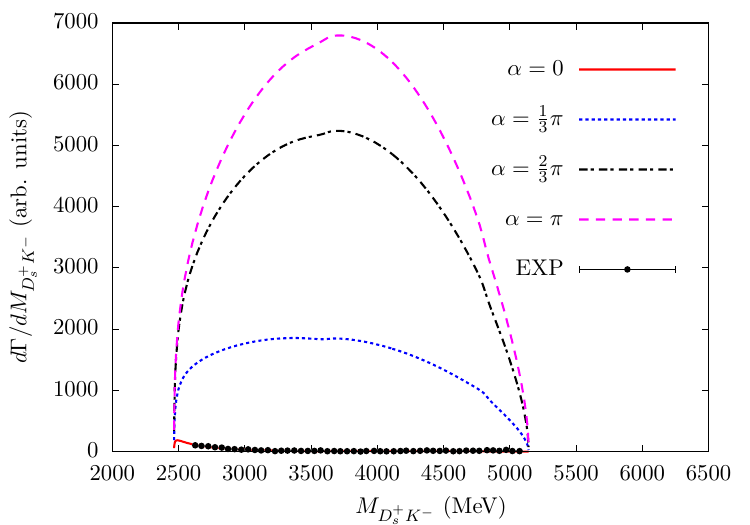}
	}	
	\subfigure[]{
		\includegraphics[scale=0.65]{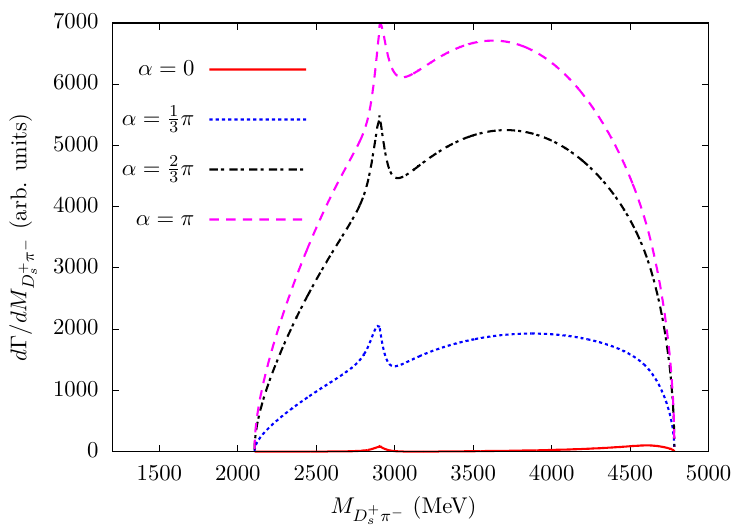}
	}
	\caption{The $D_s^+K^-$ (a) and $D_s^+\pi^-$ (b)  invariant mass distributions of the process $B^-\to D_s^+K^-\pi^-$ with the fitted parameters $\Gamma_{T_{c\bar{s}0}^0\to D_s^{+}\pi^-}=10.45$~MeV and $\phi=0.35\pi$, where the phase $\alpha$ is taken to be $\alpha=0$, $\frac{1}{3}\pi$, $\frac{2}{3}\pi$, and $\pi$, respectively. The Belle data are taken from the Ref.~\cite{Belle:2009hlu}. }\label{fig:dwidth-alpha}
\end{figure*}

\begin{figure*}[htbp]
	\subfigure[]{
		\includegraphics[scale=0.65]{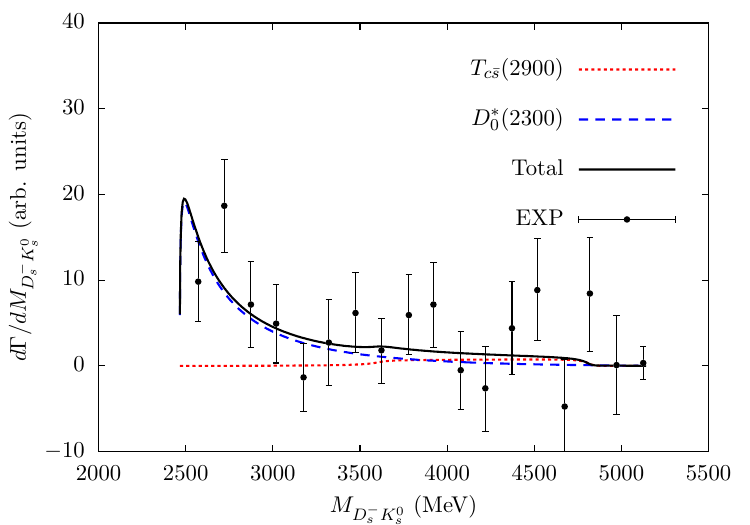}
	}	
	\subfigure[]{
		\includegraphics[scale=0.65]{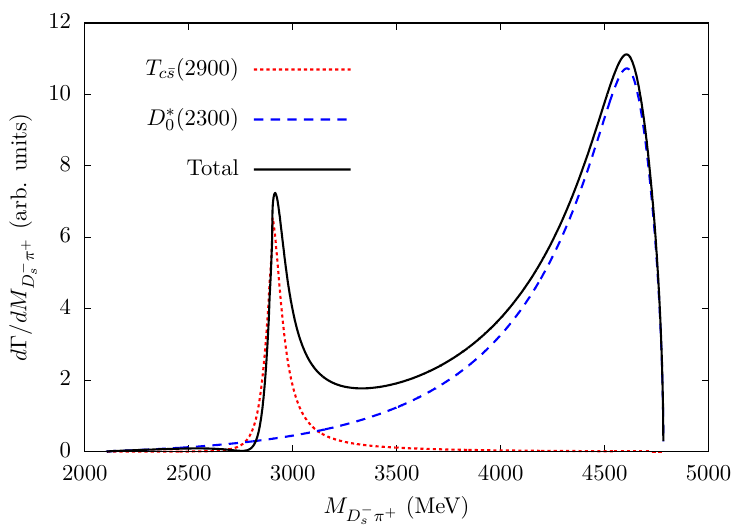}
	}
	\caption{The $D_s^-K_s^0$ (a) and $D_s^-\pi^+$ (b)  invariant mass distributions of the process $B^0\to D_s^-K_s^0\pi^+$ with the fitted parameters $\Gamma_{T_{c\bar{s}0}^0\to D_s^{+}\pi^-}=4.99$~MeV and $\phi=-0.64\pi$. The Belle data are taken from the Ref.~\cite{Belle:2014agw}. }\label{fig:dwidth-DsK0pi}
\end{figure*}


We first show the transition amplitude $t_{i\to D_s\bar{K}}$ of Eq.~(\ref{BS}) in Fig.~\ref{fig:square-T}. The red-dotted  curve shows the modulus squared of the transition amplitude $t_{D_s\bar{K}\to D_s\bar{K}}$, the blue-dashed curve shows the modulus squared of the transition amplitude $t_{D\eta\to D_s\bar{K}}$, and the black-solid curve shows the modulus squared of the transition amplitude $t_{D\pi\to D_s\bar{K}}$. One can find that the modulus squared of the transition amplitude $|t_{D\pi\to D_s\Bar{K}}|^2$ has two peaks around 2100~MeV and 2450~MeV, respectively, in consistent with the conclusion of Ref.~\cite{Montana:2020vjg}. Since the lower pole is far from the $D_s\bar{K}$ threshold, the enhancement structure near the $D_s\bar{K}$ threshold of the process $B^-\to D_s^+K^-\pi^-$ should be mainly due to the contribution from the high pole.

In our formalism, we only have one free parameter, the phase $\phi$ of Eq.~(\ref{ttotal-interference}). 
Thus, we present our results of the $D_s^+K^-$ and $D_s^+\pi^-$ invariant mass distributions with the different values of phase $\phi=0$, $\frac{1}{3}\pi$, $\frac{2}{3}\pi$, and $\pi$ in Figs.~\ref{fig:dwidth-DsK-interference} and \ref{fig:dwidth-dspi-interference}, respectively.  We also show the Belle measurements on the $D_s^+K^-$ invariant mass distribution of the $B^-\to D_s^+K^-\pi^-$ events in  Fig.~\ref{fig:dwidth-DsK-interference}, where the Belle data have been rescaled for comparison~\cite{Belle:2009hlu}\footnote{The first three data of Belle are lower than our predictions, which may be due to the lower detection efficiencies~\cite{Belle:2009hlu}, and we do not show them here.}.
One can find that, with different values of the phase $\phi$, our results of the $D_s^+K^-$ invariant mass distributions are in good agreement with the Belle measurements in the region $2600\sim4000$~MeV, and the enhancement near the threshold should be due to the resonance $D_0^*(2300)$. In Fig.~\ref{fig:dwidth-dspi-interference}, one can find a clear peak around 2900~MeV of the $D_s^+\pi^-$ invariant mass distribution, which could be associated to the $T_{c\bar{s}0}(2900)$, and the lineshape of the peak is distorted by the interference with different values of phase $\phi$.

However, in the high energy region of the $D_s^+K^-$ invariant mass distribution of Fig.~\ref{fig:dwidth-DsK-interference}, our results are smaller than the Belle measurements~\cite{Belle:2009hlu}, which implies that the contribution from the $T_{c\bar{s}0}(2900)^0$ may be underestimated. 
Thus, we take the decay width $\Gamma_{T_{c\bar{s}0}^0\to D_s^+\pi^-}$ and the phase $\phi$  to be free parameters, and fit them to the $D_s^+K^-$ invariant mass distribution of the Belle measurements~\cite{Belle:2009hlu}, and obtain the $\chi^2/N_{dof}=3.39$, and the fitted parameters $\Gamma_{T_{c\bar{s}0}^0\to D_s^+\pi^-}=(10.45\pm 1.31)$~MeV and  $\phi=(0.35\pm 0.09)\pi$, where the width $\Gamma_{T_{c\bar{s}0}^0\to D_s^+\pi^-}=10.45\pm 1.31$~MeV is close to the upper limit of the prediction of Ref.~\cite{Yue:2022mnf}.
With these fitted parameters, we have shown the $D_s^+K^-$ and $D_s^+\pi^-$ invariant mass distributions in Figs.~\ref{fig:dwidth-interference-fit}(a) and \ref{fig:dwidth-interference-fit}(b), respectively. One can find that, our results of the $D_s^+K^-$ invariant mass distribution are in good agreement with the Belle measurements in the region 2600$\sim$4800~MeV~\cite{Belle:2009hlu}, and the peak of the $T_{c\bar{s}0}(2900)^0$ in the $D_s^+\pi^-$ invariant mass distribution is more significant. Meanwhile, we also predict the Dalitz plot of ``$M_{D_s^+\pi^-}$" vs. ``$M_{D_s^+K^-}$" for the process $B^-\to D_s^+K^-\pi^-$ in Fig.~\ref{fig:Dalitz-dwidth}, and one can find that the $T_{c\bar{s}0}(2900)^0$ mainly contributes to the high energy region of the $D_s^+K^-$ invariant mass distribution. Our predictions could be tested by future measurements.

In this work, we assume the coupling constants appeared in Eqs.~(\ref{t21}) and (\ref{t22}) are real and positive. Indeed, the coupling constants could complex, thus we  multiply the Eq.~(\ref{t21}) by an interference phase factor $e^{i\phi'}$ to account for this effect. With the fitted parameters $\Gamma_{T_{c\bar{s}0}^0\to D_s^+\pi^-}=10.45$~MeV and $\phi=0.35\pi$, we have presented the $D_s^+K^-$ and $D_s^+\pi^-$ invariant mass distributions for phase $\phi'=0$, $\frac{1}{3}\pi$, $\frac{2}{3}\pi$, and $\pi$ in Figs.~\ref{fig:dwidth-phiprime}(a) and \ref{fig:dwidth-phiprime}(b), respectively. One can find that, the $D_s^+K^-$  invariant mass distribution has a minor change, and the strength of the $T_{c\bar{s}0}$ has some change. However, the most important is that, the peak position does not change, and is always very clear for different values of phase $\phi'$. 

One maybe note that the amplitude $\mathcal{T}^{D^*_0(2300)}$ of Eq.~(\ref{t-D2300}) has two terms, $\mathcal{T}^{\rm tree}$ and $\mathcal{T}^S$. Since the $G_it_{i\to D_s\bar{K}}$, involved in the term $\mathcal{T}^S$,  has included the dynamical information and is complex, the extra phase factor between $\mathcal{T}^{\rm tree}$ and $\mathcal{T}^S$ is not needed. However, in Fig.~\ref{fig:dwidth-alpha}, we also show the results of the $D^+_sK^-$ and $D^+_s\pi^-$ invariant mass distributions by multiplying $\mathcal{T}^{\rm tree}$ by an extra phase factor $e^{i\alpha}$. One could find the lineshapes of the $D^+_sK^-$ invariant mass distribution with non-zero $\alpha$ are significantly different with the experimental data, which implies that one donot need to consider the extra phase factor between $\mathcal{T}^{\rm tree}$ and $\mathcal{T}^S$.

In Ref.~\cite{Belle:2014agw}, the Belle Collaboration has reported the $D^-_sK_s^0$ invariant mass distribution of the process $B^0 \to D^-_sK^0_s\pi^+$. With the same formalism as given in this work, we could determine the corresponding $\mathcal{Q}^2=1.59\times10^{-13}$ and $\mathcal{Q}^{\prime 2}=1.02\times10^{-12}$ with the branching fractions $\mathcal{B}(B^0\to D^*K^*K)=(1.29 \pm0.33 )\times 10^{-3}$~\cite{ParticleDataGroup:2022pth} and 
$\mathcal{B}(B^0\to D^-_sK^0_s\pi^+)=(0.47\pm0.06\pm0.05)\times 10^{-4}$~\cite{Belle:2014agw}. Furthermore, we obtain the $\chi^2/N_{dof}=0.95$, and the $\Gamma_{T_{c\bar{s}0}^0\to D_s^{+}\pi^-}=(4.99\pm 0.01)$~MeV and $\phi=(-0.64\pm 0.61)\pi$ by fitting to the Belle data. The fitted width is in agreement with the result of Ref.~\cite{Yue:2022mnf}, and we show the $D^-_sK^0_s$ and $D^-_s\pi^+$ invariant mass distributions in Fig.~\ref{fig:dwidth-DsK0pi}. One can find that our prediction of the $D^-_sK^0_s$ invariant mass distribution are in good agreement with the Belle data~\cite{Belle:2014agw}, and one peak around 2900~MeV is expected to be observed in the $D^-_s\pi^+$ invariant mass distribution.


\section{Conclusions }
Recently, the LHCb Collaboration has  reported their amplitude analysis of the processes $B^0\to \bar{D}^0D_s^+\pi^-$ and $B^+\to D^-D_s^+\pi^+$, where two states $T_{c\bar{s}0}(2900)^0$ and $T_{c\bar{s}0}(2900)^{++}$ were observed in the $D_s\pi$ invariant mass distributions.  The resonance parameters of these two resonances indicate that they are two of the isospin triplet.
Motivated by those observations of the LHCb, we propose to search for the  state $T_{c\bar{s}0}(2900)^{0}$ in the process $B^-\to D_s^+K^-\pi^-$.

In the picture of $T_{c\bar{s}0}(2900)$ as a $D^{*}K^{*}$ molecular state, we have investigated the process $B^-\to D_s^+K^-\pi^-$ by taking into account the $S$-wave $D_s^*\rho$ and $D^*K^*$ interactions, and the $S$-wave pseudoscalar meson-pseudoscalar meson interactions, which dynamically generate the resonance $D_0^*(2300)$. We have found that there is a near-threshold enhancement in the $D_s^+K^-$ invariant mass distribution, which is in good agreement with the Belle measurements. Indeed, this enhancement structure is mainly due to the high pole of the $D_0^*(2300)$. In addition, a clear peak structure appears around 2900~MeV in the $D_s^+\pi^-$ invariant mass distribution, which should be associated to the $T_{c\bar{s}0}(2900)$.

Considering that our predictions for the $D_s^+K^-$ invariant mass distribution are lower than the Belle measurements in the high energy region, we take the decay width $\Gamma_{T_{c\bar{s}0}^0 \to D_s^{+}\pi^-}$ and the phase between two amplitudes $\phi$ to be free parameters, and obtain $\Gamma_{T_{c\bar{s}0}^0\to D_s^{+}\pi^-}=(10.45\pm 1.31)$~MeV and $\phi=(0.35\pm 0.09)\pi$ by fitting to Belle measurements. Our new results show a more significant peak of $T_{c\bar{s}0}(2900)$ in the $D_s^+\pi^-$ invariant mass distribution.
Furthermore, we have also discussed the effects of the interference phase between the coupling constants.

With the formalism presented in this work, we have also analyzed the Belle measurements about the process  $B^0 \to D^-_sK^0_s\pi^+$. With the fitted parameters $\Gamma_{T_{c\bar{s}0}^0\to D_s^{+}\pi^-}=(4.99\pm 0.01)$~MeV and $\phi=(-0.64\pm 0.61)\pi$, our prediction of the $D^-_sK^0_s$ invariant mass distribution are in agreement with the Belle measurements, and one peak around 2900~MeV is expected to be observed in the $D^-_s\pi^+$ invariant mass distribution.

In summary, within some theoretical approximation, our results of the $D_s^+K^-$ invariant mass distribution could well reproduce near-threshold enhancement structure observed by Belle Collaboration, and the predictions of the $T_{c\bar{s}}(2900)$ peak in the $D_s^+\pi^-$ could be tested by the LHCb and Belle~II experiments in future. The more precise measurements of the process $B^-\to D_s^+K^-\pi^-$ would shed light on the nature of the $T_{c\bar{s}0}(2900)^0$ resonance.

\section*{Acknowledgements}

This work is supported by the National Natural Science Foundation of China under Grant Nos. 11775050, 12335001, 12175037, and 12192263, the Natural Science Foundation of Henan under Grant Nos. 222300420554, 232300421140, the Project of Youth Backbone Teachers of Colleges and Universities of Henan Province (2020GGJS017), and the Open Project of Guangxi Key Laboratory of Nuclear Physics and Nuclear Technology, No. NLK2021-08.

\end{document}